\documentclass[aps,pre,twoside,twocolumn]{revtex4}
\usepackage{graphicx}
\usepackage{amssymb}

\newcommand{\br}{{\bf r}}

\newcommand{\cH}{{\cal H}}

\newcommand{\half}{\frac{1}{2}}
\newcommand{\f}{\frac}

\newcommand{\const}{{\mathrm{const.}}}
\newcommand{\la}{\langle}
\newcommand{\ra}{\rangle}

\begin{document}

\title{Virus shapes and buckling transitions in spherical shells}

\author{Jack Lidmar}
\affiliation{
Department of Physics,
Royal Institute of Technology, AlbaNova,
SE-106 91 Stockholm, Sweden
}

\author{Leonid Mirny}
\affiliation{
Harvard-MIT Division of Health Science and Technology,
Massachusetts Institute of Technology,
Cambridge, MA 02139,
USA
}

\author{David R. Nelson}
\affiliation{
Lyman Laboratory of Physics,
Harvard University,
Cambridge, MA 02138,
USA
}

%% \date{\today}
\date{June 27, 2003}

\def\calP{{\cal P}}
\def\calL{{\cal L}}
\def\cH{{\cal H}}

\begin{abstract}
We show that the icosahedral packings of protein capsomeres proposed
by Caspar and Klug for spherical viruses become unstable to faceting
for sufficiently large virus size, in analogy with the buckling
instability of disclinations in two-dimensional crystals. Our model,
based on the nonlinear physics of thin elastic shells, produces
excellent one parameter fits in real space to the full
three-dimensional shape of large spherical viruses. The faceted shape
depends only on the dimensionless Foppl-von K{\'a}rm{\'a}n number
$\gamma=YR^2/\kappa$, where $Y$ is the two-dimensional Young's modulus
of the protein shell, $\kappa$ is its bending rigidity and $R$ is the
mean virus radius. The shape can be parameterized more quantitatively
in terms of a spherical harmonic expansion.
We also investigate elastic shell theory for extremely large $\gamma$,
$10^3<\gamma<10^8$, and find results applicable to icosahedral shapes
of large vesicles studied with freeze fracture and electron
microscopy.
\end{abstract}

\pacs{}

\maketitle

\section{Introduction}					\label{sec:intro}

Understanding virus structures is a rich and challenging
problem~\cite{Chiu}, with a wealth of new information now becoming
available.  Although traditional X-ray crystallography still allows
the most detailed analysis~\cite{Harrison-91}, three-dimensional
reconstructions of icosahedral viruses from cryo-electron micrographs
are now becoming routine~\cite{Baker}.  Electron microscope images of
many identical viruses in a variety of orientations are used to
reconstruct a three-dimensional image on a computer, similar to CT
(computed tomography) scans in medical imaging. There are now, in
addition, beautiful single molecule experiments which measure the work
needed to load a virus (bacteriophage $\phi$29) with its DNA
package~\cite{Smith}. The aim of this paper is to explore the elastic
parameters and physical ideas which determine the shapes of viruses
with an icosahedral symmetry, using the theory of thin elastic
shells~\cite{Landau-Lifshitz}.

\begin{figure}
\centerline{\includegraphics{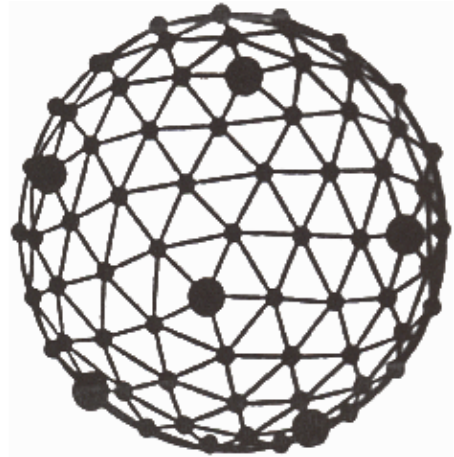}}
\caption{
\label{fig:1}
A right-handed (3,1) triangulated net (icosadeltahedron) used to
describe virus structure. The (1,3) structure is left-handed. }
\end{figure}

The analysis of approximately spherical viruses dates back to
pioneering work by Crick and Watson in 1956~\cite{Crick}, who argued
that the small size of the viral genome requires identical structural
units packed together with an icosahedral symmetry.  These principles
were put on a firm basis by Caspar and Klug in
1962~\cite{Caspar-Klug}, who showed how the proteins in a viral shell
(the ``capsid'') could be viewed as icosadeltahedral triangulations of
the sphere by a set of pentavalent and hexavalent morphological units
(``capsomers'').  The viral shells (there can also be an outer
envelope composed of additional proteins and membrane elements from
the host cell) are characterized by a pair of integers $(h,k)$ such
that the number of morphological units is $N = 10(h^2 + hk + k^2) +
2$. To get from one pentavalent capsomer to another, one moves $h$
capsomers along a row of near-neighbor bonds on the sphere, turns 120
degrees and moves another $k$ steps. Euler's theorem relating the
number of vertices, edges and faces of a spherical triangulation
insures that the number of capsomers in five-fold environments is
exactly 12~\cite{Coxeter}.  A simple icosahedron of 12 morphological
units corresponds to (1,0) while soccer balls and C$_{60}$ fullerene
molecules are (1,1) structures with $N=32$ polygons. A (3,1)
icosadeltahedron $N=132$ is shown in Fig.~\ref{fig:1}. The polyoma
virus (SV40) is a (2,1) structure with 72 capsomeres, while the much
larger adenovirus and herpes simplex virus are (5,0) and (4,0)
structures with 252 and 162 morphological units, respectively.
Structures like that in Fig.~\ref{fig:1} with $h$ and $k$ nonzero and
$h \not= k$ are chiral.

%Fig 1

\begin{figure}
\centerline{\includegraphics[width=\linewidth]{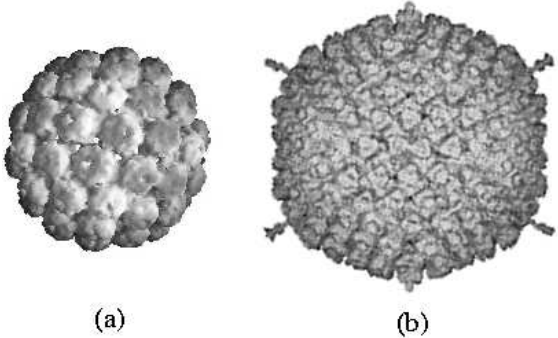}}
\caption{
\label{fig:2}
The polyoma virus (a) is approximately spherical, while the larger
adenovirus (b) is more faceted (not to scale). Images from
Ref.~\cite{Reddy}.}
\end{figure}

Note that the relatively small polyoma virus (diameter 440 Angstroms)
is round (see Fig.~\ref{fig:2}a), while the much larger herpes simplex
virus (diameter 1450 Angstroms) has a more angular or faceted
shape~\cite{Reddy} (see Fig.~\ref{fig:2}b).  Faceting of large viruses
is in fact a common phenomenon; the protein subunits of different
viruses, moreover, are very similar (see below).
If these protein assemblies are characterized by elastic constants and
a bending rigidity~\cite{Landau-Lifshitz}, we can ask how deviations
from a spherical shape develop with increasing virus radius, which
scales roughly as the square root of the number of morphological
units.

%Fig 2

In support of the idea that viruses with different overall capsid size
are composed of nearly identical monomers, we note that most viral
coat and capsid proteins have about the same size, molecular weight,
amino acid composition and, most importantly, the same folded
structure in three dimensions~\cite{Murzin}.  It is known, moreover,
that protein structure determines the mechanical properties of
proteins~\cite{Carrion-Vazquez,Kellermayer}.  Hence, the similarity of
the protein structure of the coat/capsid proteins suggests similar
mechanical properties.  In addition, the presence of the same fold of
capsid proteins in unrelated viruses (bacterial phages, plant viruses,
insect viruses and animal viruses) indicate that the fold and its
mechanical properties are conserved in evolution and could be
essential for proper virus assembly.

\begin{figure}
\centerline{\includegraphics[width=\linewidth]{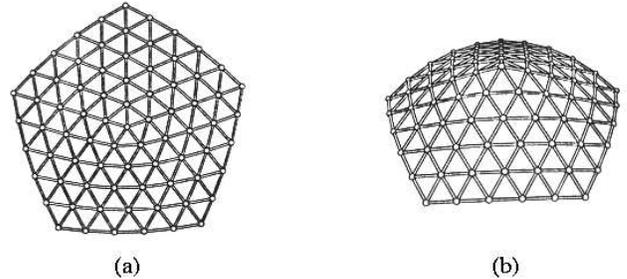}}
\caption{
\label{fig:3}
The fivefold disclination in a triangular lattice. In viruses, the
points would correspond to capsomeres, while in lecithin vesicles they
correspond to lipid molecules.  The highly strained flat space
configuration is shown in (a). The buckled form which arises for
$\gamma \geq $ 154 is shown in (b).}
\end{figure}

In this paper, we argue that
the faceting of large viruses is caused by a buckling
transition associated with the 12 isolated points of 5-fold symmetry.
These singularities can be viewed as disclinations in an otherwise
6-coordinated medium.  It is well known that the large strains
associated with an isolated disclination in a {\it flat} disk spanned
by a triangular lattice leads to buckling into a conical shape
for~\cite{Seung-Nelson,Nelson-Peliti}
\begin{equation}					\label{eq:1}
YR^2/\kappa\geq 154,
\end{equation}
where $Y$ is the two-dimensional Young's modulus, $\kappa$ is the
bending rigidity and $R$ is the disk radius. The energy of a single
5-fold disclination with ``charge'' $s = 2\pi/6$ centered in a flat
array of proteins of size $R$ is approximately
\begin{equation}					\label{eq:2}
E_5\approx {1\over 32\pi} s^2YR^2.
\end{equation}
However, above a critical buckling radius
$R_b\approx\sqrt{154\kappa/Y}$, is there a conical deformation (see
Fig.~\ref{fig:3}) such that the disclination energy now only grows
logarithmically with $R$
\begin{equation}
E_5\approx (\pi/3)  \kappa \ln 
(R/R_b)+ {1\over 32\pi} s^2YR_b^2.
\end{equation}
One might expect a similar phenomenon for 12 disclinations confined to
a surface with a \emph{spherical} topology.  Indeed, the elastic
energy for 12 disclinations on an undeformed sphere of radius $R$ is
has a form similar to Eq.~(\ref{eq:2}), namely~\cite{Bowick}
\begin{equation}
E \approx  0.604(s^2 Y R^2/ 4\pi),
\end{equation}
where the sphere radius $R$ now plays the role of the system size.
Although it seems highly likely that these 12 disclinations can lower
their energy by buckling for large $R$, the nonlinear nature of the
underlying elastic theory~\cite{Landau-Lifshitz} leads to complex
interactions between the resulting conical deformations. A boundary
layer analysis of the von K{\'a}rm{\'a}n equations for bent plates
predicts anomalous scaling for the mean curvature in the vicinity of
the ridges connecting conical
singularities~\cite{Lobkovsky,Lobkovskywitten}.  Interesting scaling
behavior also arises in the vicinity of the apexes of the cones
themselves~\cite{Cerda}.  Another interesting physical realization of
the buckling problem lies in the faceting of lecithin vesicles at
temperatures sufficiently low so that the lipid constituents have
crystallized~\cite{Blaurock,Zemb}.

%Fig 3

In this paper, we study the ground states of crystalline particle
arrays with 12 disclinations in a spherical geometry. We find that
there are indeed striking manifestations of the buckling transition
even in the curved geometry of viral capsids or crystalline
vesicles. The nonlinear Foppl-von K{\'a}rm{\'a}n equations for thin
shells with elasticity and a bending energy are solved using a
floating mesh discretization developed and studied extensively in the
context of ``tethered surface'' models of polymerized
membranes~\cite{Kantor-Nelson}.  By taking the nodes of the mesh to
coincide with the capsomers, even small viruses can be handled in this
way, although any buckling transition will surely be smeared out
unless $R_b/a$ is large, where $a$ is the spacing between these
morphological units.  Ideas from continuum elastic theory will, of
course, be most applicable for vesicles composed of many lipids and
for large viruses --- Viruses with as many as 1692 morphological units
have been reported~\cite{Yan}.

There may be inherent limitations on the size of viral capsids that
follow from the elastic properties of thin shells.  Because larger
viruses can accommodate more genetic material, large sizes could
confer an evolutionary advantage. If, however, large viruses buckle
away from a spherical shape, the resistance of the capsid to
mechanical deformation can degrade.  As we shall see, a theory of
buckled crystalline order on spheres also allows estimates of
important macroscopic elastic parameters of the capsid shell from
structural data on the shape anisotropy.  Estimates of quantities such
as the bending rigidity and Young's modulus of a empty viral shell
might allow an understanding of deformations due to loading with DNA
or RNA \cite{Smith}.  Although some aspects of virus structure may be
accounted for by the physics of shell theory, we should emphasize that
other features could be driven by the need for cell recognition,
avoidance of immune response, etc.

\begin{figure}
\centerline{\includegraphics[width=\linewidth]{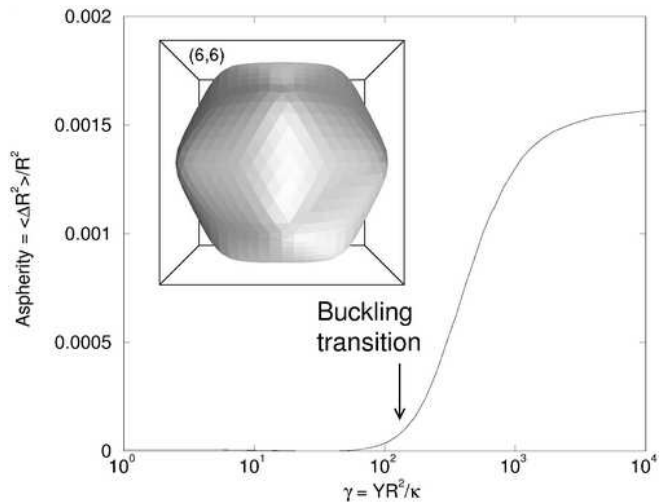}}
\caption{
\label{fig4}
Mean square asphericity as a function of $YR^2/\kappa$ for many
different icosahedral shells. The inset shows a (6,6) structure with
Foppl-von K{\'a}rm{\'a}n number $\gamma = YR^2/\kappa \approx 400$.
The arrow marks the location of the buckling transition in flat
space.}
\end{figure}

A summary of our investigations of buckling transitions in a spherical
geometry (discussed in detail in Sec.~II) is shown in Fig.~\ref{fig4}.
As illustrated in Fig.~\ref{fig4} (See also Fig.~\ref{fig:shapes})
icosahedral shells do indeed become aspherical as the ``Foppl-von
K{\'a}rm{\'a}n number'' $\gamma=YR^2/\kappa$ increases from values of
order unity to $YR^2/\kappa\gg 1$.  The mean square ``asphericity''
(deviation from a perfect spherical shape) departs significantly from
zero when $YR^2/\kappa$ exceeds 154, the location of the buckling
transition in flat space~\cite{Seung-Nelson}.  Fits of buckled viruses
or crystalline vesicles to this universal curve would allow an
experimental determination of the ratio $Y/\kappa$.  More quantitative
information on the buckled shape can be obtained by expanding the
radius $R(\theta,\phi)$ in spherical harmonics,
\begin{equation}
R(\theta,\phi)=\sum_{\ell=0}^\infty
\sum_{m=-\ell}^\ell
Q_{\ell m}
Y_{\ell m}
(\theta,\phi),
\end{equation}
and studying the rotationally invariant quadratic invariants allowed
for viruses or vesicles with icosahedral symmetry, namely
\begin{equation}
\hat{Q}_\ell=
\sqrt{\frac{1}{2\ell+1}
\sum_{m=-\ell}^\ell|Q_{\ell m}|^2} /Q_{00}
\end{equation}
with $\ell= 0, 6, 10, 12, 16, 18, \cdots$ \cite{Steinhardt}.  Although
{\it any} parameter set of the form $\{\hat{Q}_6, \hat{Q}_{10},
\hat{Q}_{12}\cdots\}$ could be consistent with an icosahedral
symmetry, all buckled objects describable by the theory of elastic
shells in fact lie on a universal curve parametrized by the value of
$YR^2/\kappa$.  Deviations from this curve would presumably describe
biological features such as the protrusions of the adenovirus in
Fig.~\ref{fig:2}.

%Fig 4

In Sec.~II, we describe our theoretical results for disclination
buckling in the icosadeltahedral spherical shells proposed by Caspar
and Klug as models of viruses~\cite{Caspar-Klug}. The energy,
mean-square aspherity, and spherical harmonic content of these shells
are determined as a function of the Foppl-von K{\'a}rm{\'a}n number,
discussed above,
\begin{equation}
\gamma=YR^2/\kappa.
\end{equation}
Most viruses have either Foppl-von K{\'a}rm{\'a}n number $\gamma
\lesssim 150$ (implying a close to spherical shape) or $200 \lesssim
\gamma \lesssim 1500$ (noticeably buckled).  Higher von
K{\'a}rm{\'a}n numbers describing objects with very sharp corners
cannot be obtained for viruses with $R \lesssim 0.2 \mu$m composed of
finite size proteins.  Of course, very high von K{\'a}rm{\'a}n numbers
{\em are} possible for spherical vesicles with crystalline order
composed of much smaller lipid molecules~\cite{Blaurock,Zemb}.

We have studied the scaling of the curvature $C$ at the creases
formed after the shells buckle for large $\gamma$. We eventually
recover the scaling proposed and studied in other geometries by
Lobkovsky, Witten and collaborators~\cite{Lobkovsky,Lobkovskywitten},
but only for very large $\gamma$, $\gamma\ge 10^7$,
appropriate for the buckled icosahedral vesicles described in
Ref.~\cite{Zemb}.

In Sec.~III, we discuss briefly the relevance of our work to
icosahedral viruses from the library of those whose structures have
been determined by diffraction methods or by cryro-electron
microscopy~\cite{Chiu,Harrison-91,Baker,Reddy}.  For viruses large
enough to buckle, the model seems to account well for the deviation
from the spherical shape using the single adjustable parameter
$\gamma$. These fits in turn provide information about the ratio
$Y/\kappa$ of the Young's modulus to the bending rigidity.
We also discuss the possible relevance of spontaneous curvature terms
and outward pressure induced by packaged DNA and RNA in this Section.
See Ref.~\cite{Seifert} for a discussion of similar issues for the
shapes of liquid membranes with a spherical topology.

\section{Defects on Curved Surfaces}
\label{sec:II}

Topological defects play a very important role in crystalline matter.
A particularly common type of defect, the dislocation, is largely
responsible for the strength of materials, and in two-dimensional
systems the unbinding of dislocations may drive the crystal into a
hexatic phase~\cite{Nelson-book}.  Disclinations on the other hand,
are much less common in the crystalline phase because of their very
large energy.  In quasi-two-dimensional curved surfaces, however, the
situation may be quite different.  The Gaussian curvature of the
surface will ``screen'' out the strain around the defect and thus
lower the energy.  Moreover, when a crystalline surface is bent to
form a closed surface with spherical topology, defects are necessarily
introduced into the lattice.  For a triangular lattice on a sphere the
number of disclinations has to be at least 12.  More generally, the
difference $N_5-N_7$ between the number of fivefold disclinations and
the number of sevenfold disclinations (assuming defects with
coordination numbers other than 5, 6 or 7 are absent) is precisely 6
times the Euler characteristic $\chi$ of the triangulated surface,
i.e., $N_5-N_7=6\chi=12(1-g)$, where $g$ is the genus or number of
handles.  Thus for shapes such that $g= 0$ (the sphere) and torii with
extra handles $(g\ge 2)$, the ground state must necessarily contain
disclinations. (A simple torus has genus $g=1$, so there is no
topological necessity for defects.)

In this section we study closed surfaces of spherical topology.  The
12 disclinations present in the crystalline lattice structure can be
expected to dominate the energetics and affect the overall shape of
the structure.  The repulsive interaction between the 12 disclinations
will favor an arrangement which maximizes their separation. Absent an
instability toward grain boundaries~\cite{Bowick}, this leads to a
configuration with icosahedral symmetry, where the disclinations sit
at the vertices.  We show below that, as a result of competition
between strain and bending energies, these structures may undergo a
buckling-like transition (smeared by finite size effects) from a
smooth round shape to a sharply faceted shape as the size or elastic
constants are varied.

\subsection{Disclination Buckling on Spheres}
\label{sec:II-A}

We assume a thin shell described by a continuum elastic theory, with
energy $\cH = \cH_s + \cH_b$~\cite{Landau-Lifshitz}, including both an
in-plane stretching energy,
\begin{equation}					\label{stretch}
	\cH_{s} = \half\int dS \left( 2 \mu u_{ij}^2 + \lambda
			u_{kk}^2 \right),
\end{equation}
where $u_{ij}$ is the strain tensor, $\mu$ and $\lambda$ are the 2D
Lam{\'e} coefficients, and a bending energy,
\begin{equation}					\label{bend}
	\cH_{b} = \half\int dS \left( \kappa H^2 + 2 \kappa_G K \right),
\end{equation}
where $\kappa$ is the bending rigidity, $\kappa_G$ the Gaussian
rigidity, $H$ and $K$ the mean and Gaussian curvatures, respectively.
(If $R_1$ and $R_2$ are the principal radii of curvature, $H=1/R_1 +
1/R_2$ and $K = 1/R_1 R_2$~\cite{Landau-Lifshitz}.)  For a closed
surface with fixed topology the Gaussian curvature integrates to a
constant (provided that $\kappa_G$ is constant) and will henceforth be
dropped, as it will have no influence on the shape.  Instead of the
Lam{\'e} coefficients we will use the 2D Young's modulus $Y$ and the
Poisson ratio $\nu$, which are given by
\begin{equation}
	Y   = \f{4 \mu (\mu + \lambda)}{2 \mu + \lambda}, \qquad
	\nu = \f{\lambda}{2 \mu + \lambda}.
\end{equation}
Taking the variation of $\cH$ with respect to coordinates
parameterizing the surface leads to the Foppl-von K{\'a}rm{\'a}n
equations which are highly nonlinear and whose solution even in simple
geometries is very difficult~\cite{Landau-Lifshitz}.

The energy of disclinations on flexible crystalline membranes was
studied by Seung and Nelson in Ref.~\cite{Seung-Nelson}.  As discussed
in the Introduction, for a thin flat plate of finite radius $R$ with
an isolated 5-fold disclination at the center, the energy grows
quadratically with the radius, $E_{\mathrm{flat}} \simeq A Y R^2$,
where $A\approx \pi/288$ is a numerical constant.  If the disclination
is allowed to buckle out of the plane this energy is reduced, and
grows logarithmically for large $R$ with a coefficient proportional to
$\kappa$.  In the inextensional limit $Y\to\infty$ the problem
simplifies considerably and the energy is $E_{\mathrm{buckled}} \simeq
B \kappa \ln\left(R/a\right)$, where $B \approx \pi/3$ and $a$ the
lattice constant.  Thus, for small plates the flat solution is lower
in energy, whereas for large plates the buckled solution wins.  An
instability separating these two regimes occurs when
$E_{\mathrm{flat}} \approx E_{\mathrm{buckled}}$, or $YR^2/\kappa
\approx B/A$.  A detailed calculation in Ref.~\cite{Seung-Nelson}
found the transition at a critical value of $YR^2/\kappa \approx 154$.

The spherical shapes of icosahedral symmetry we are interested in here
can approximately be thought of as being composed of 12 disclinations,
and should therefore undergo a similar transition from flat to
buckled.  Because the surface of the sphere is already curved, and
hence breaks the up-down symmetry that was present for a thin plate,
it is not clear that a sharp instability survives in this case.
However, even if this is the case, we might still expect to see
remnants of the transition in the form of a sharp if not singular
crossover.  We construct below simple estimates of the energies and
transitions involved.

The total energy of the closed shell in the vicinity of the transition
is approximately 12 times the energy of a disclination (with radius
approximately equal to the radius of the sphere), plus contributions
from the background curvature of a sphere, given approximately by
$8\pi \kappa + 4\pi \kappa_G$.  In the inextensional limit the short
distance cutoff in the buckled disclination energy was provided by the
lattice constant.  For finite $Y$ the cutoff is determined instead by
a balance of strain and bending.  We may in this case approximate the
buckled disclination by a flat inner region $r<R_b$ with energy
$AYR_b^2$ and an outer buckled region $r>R_b$ with energy $B \kappa
\ln\left(R/R_b\right)$.  Minimizing the sum with respect to $R_b$
gives $Y R_b^2/\kappa = B/2A$, and $E \simeq \kappa B \left(\half +
\ln\left(R/R_b\right)\right)$.

Because the disclination energy is independent of the two-dimensional
Poisson ratio and the Gaussian rigidity drops out, the solution
depends only on a single dimensionless parameter, $\gamma =
YR^2/\kappa$, which we term the Foppl-von K{\'a}rm{\'a}n number.  Note
that, if the 2D elastic theory derives from a thin shell of finite
thickness $d$ built from a 3D isotropic elastic medium, we have
$\gamma = 12 \left(1 - \nu_3^2\right) \left( R / d \right)^2$, where
$\nu_3$ is the three-dimensional Poisson ratio, a result independent
of the 3D Young's modulus $Y_3$~\cite{Landau-Lifshitz}.  In summary,
we then expect the energy of the closed shell with twelve
disclinations to approximately be
\begin{equation}				\label{energy}
 \f{E}{\kappa} \simeq \left\{
  \begin{array}{ll}
    6B \gamma/\gamma_b + D, &
	\gamma < \gamma_b \\
    6B \left(1 + \ln\left(\gamma/\gamma_b\right)\right) + D, &
	\gamma > \gamma_b \\
  \end{array}
 \right.,
\end{equation}
where $\gamma_b = YR_b^2/\kappa$.  The background curvature gives a
constant contribution $D \approx 4\pi\left(2+\kappa_G/\kappa\right)$,
leading as well to a small shift in the disclination energies and
hence in $A$ and $B$.  For a perfect sphere, e.g., $12A \approx 0.604
\pi/36$~\cite{Bowick}.

As the disclinations become more sharply buckled the whole structure
will become more faceted.  This leads to the formation of ridges
connecting the vertices of the icosahedron.  The energy of similar
ridges has been studied recently~\cite{Lobkovsky,Lobkovskywitten},
uncovering some remarkable scaling relations.  As the ridges become
sharper upon increasing $\gamma$, the energy will increase as
$E_{\mathrm{ridge}}/\kappa \simeq 1.24 \alpha^{7/3} \left(Y L^2 /
\kappa\right)^{1/6}$, where $\alpha$ is the angle in radians and $L$
is the length of the ridge.  In this regime the shape is very close to
an icosahedron with sharp facets, with $\alpha \approx 0.365$ and $L
\approx 1.23 R$, and a total of 30 ridges.  In the limit of very large
$\gamma = Y R^2/\kappa$ the energy should therefore crossover to
\begin{equation}			\label{energy-largegamma}
  E /\kappa \simeq C \gamma^{1/6} + \const, \quad \gamma \to \infty,
\end{equation}
where $C \approx 3.8$.

To check these arguments and to calculate more properties of the
shells we now present numerical calculations.

%Fig 5
\begin{figure}
\centerline{\includegraphics[width=\linewidth]{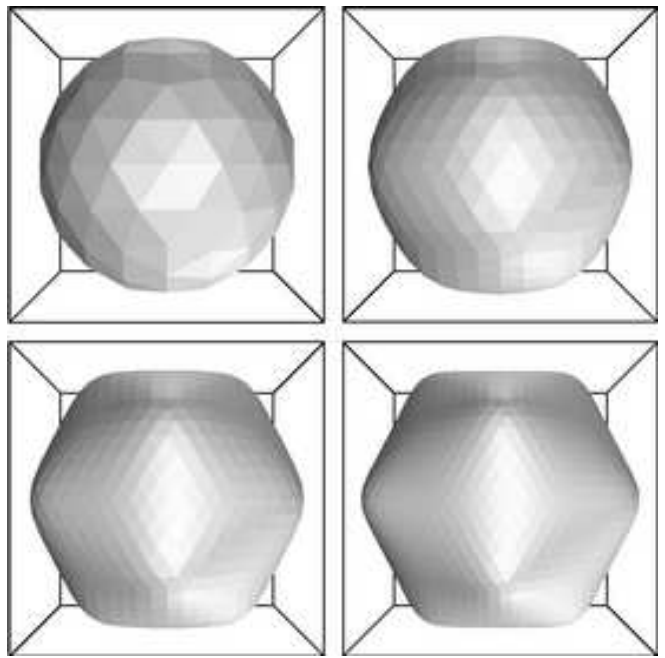}}
\caption{\label{fig:shapes}
Numerically calculated shapes with $(h,k)$ indices (2,2), (4,4),
(6,6), and (8,8) for fixed $\tilde{\kappa}$ = 0.25, and fixed spring
constant $\epsilon = 1$.  The Foppl-von K{\'a}rm{\'a}n numbers for
these shapes are $\gamma \approx 45$, $176$, $393$ and $694$.
}
\end{figure}

\subsection{Numerical Results}
\label{sec:II-B}

For numerical calculations it is useful to consider discretized
versions of Eqs.~(\ref{stretch}) and (\ref{bend})~\cite{Seung-Nelson}:
\begin{equation}
	\cH_{s} = \f{\epsilon}{2} \sum_{\left<ij\right>}
		  \left( \left| \br_i - \br_j \right| - a \right)^2
\end{equation}
and
\begin{equation}
	\cH_{b} = \f{\tilde\kappa}{2} \sum_{\left<IJ\right>}
		  \left( {\bf \hat n}_I - {\bf \hat n}_J \right)^2.
\end{equation}
Here $\left<ij\right>$ denote pairs of nearest neighbor vertices
(which we identify with the centers of the capsomers of a virus), with
positions $\br_i$, and $\left<IJ\right>$ pairs of nearest neighbor
plaquettes of a triangulated surface, with unit normals ${\bf \hat
n}_I$.  In the continuum limit this model becomes equivalent to
Eqs.~(\ref{stretch}) and (\ref{bend}) with
parameters~\cite{Seung-Nelson}
\begin{eqnarray}
	Y = \f{2}{\sqrt{3}}\epsilon, \quad && \nu=\f{1}{3} \\
	\kappa = \f{\sqrt{3}}{2}\tilde{\kappa}, \quad &&
	\kappa_G=-\f{4}{3}\kappa
\end{eqnarray}
The relation $\kappa_G = -4\kappa/3$ was calculated by comparing the
bending energy of a triangulated cylinder and sphere with the
corresponding continuum expressions, and differs from the one used in
Ref.~\cite{Seung-Nelson}.
Closed triangular surfaces of icosahedral symmetry are constructed for
non-negative integers $(h,k)$ according to the geometric principles of
Caspar and Klug~\cite{Caspar-Klug}, and the minimum energy
configuration is found numerically using a conjugate gradient method
for different values of $\tilde \kappa$.  As discussed in the
Introduction, the integers $(h,k)$ denote the number of steps along
the two lattice vectors between two neighboring disclinations of the
structure.
Figure~\ref{fig:shapes} shows some examples of the resulting shapes.
Shapes of varying size with ``$T$-numbers'' as large as $T = h^2 + hk +
k^2 \approx 1500$ are studied.  Below we calculate some properties of
the shells to characterize the shapes quantitatively.

\subsubsection{Energy}

\label{sec:energy}

\begin{figure}
\centerline{\includegraphics[width=\linewidth]{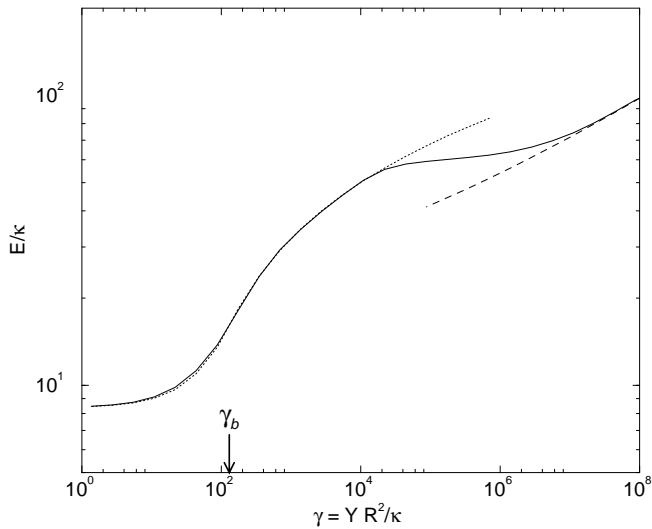}}
\caption{
\label{fig:energy}
Total energy. Dotted and dashed lines are fits to Eq.~(\ref{energy})
and (\ref{energy-largegamma}), respectively.  The arrow indicates the
value of $\gamma_b$ obtained from the fit.  }
\end{figure}

We first plot the total energy as a function of the Foppl-von
K{\'a}rm{\'a}n number $\gamma$ in Fig.~\ref{fig:energy}.  
Similar to results for an isolated disclination in a disk with free
boundary conditions~\cite{Seung-Nelson}, the energy crosses over from
a ``flat'' regime dominated by stretching energy to a ``buckled''
regime dominated by bending energies.  Fitting the functional form
Eq.~(\ref{energy}) gives $A \approx 0.005$, $B \approx 1.30$,
$\gamma_b \approx 130$, which compares quite well with the estimates
above.
For large $\gamma \gtrsim 10^7$ a crossover to the form given by
Eq.~(\ref{energy-largegamma}) (indicated by the dashed line in the
figure) occurs.

%Fig 6

\subsubsection{Aspherity}
\label{sec:aspherity}

\begin{figure}
\centerline{\includegraphics[width=\linewidth]{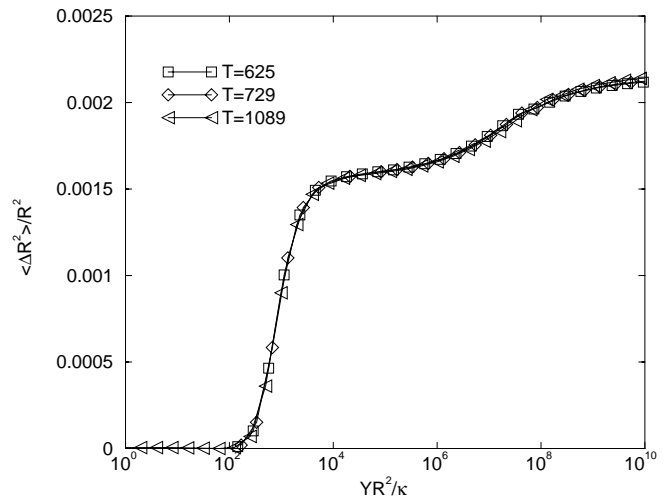}}
\caption{
\label{fig:aspherity}
Mean squared aspherity (see Eq.~(\ref{aspherity})) as a function of
$\gamma = YR^2/\kappa$ for various ``triangulation numbers'' $T=h^2 +
hk + k^2$.  Data from shapes of three different sizes collapse onto a
single universal curve, indicating that the continuum limit has been
reached for these sizes.  }
\end{figure}

As a measure of the deviation from a perfectly spherical shape
centered on the origin we calculate the mean squared aspherity,
defined by
\begin{equation}					\label{aspherity}
	\f{\left<\Delta R^2\right>}{\la R \ra^2} =
	\f{1}{N} \sum_{i=1}^N \f{(R_i - \la R\ra)^2}{\la R\ra^2},
\end{equation}
where $R_i$ is the radial distance of vertex $i$ and $\la R \ra$ is
the mean radius,
\begin{equation}
	\la R\ra = \f{1}{N} \sum_{i=1}^N R_i.
\end{equation}
The result, which is displayed in Fig.~\ref{fig:aspherity}, shows a
rather sharp, but non-singular crossover from spherical shape to
faceted at roughly $\gamma = Y R^2/\kappa \approx 150$.  The second
increase around $Y R^2/\kappa \approx 10^7$ coincides roughly with the
sharpening of the ridges, where the asymptotic scaling in
Eq.~(\ref{energy-largegamma}) sets in.
Note that the log-linear plot of Fig.~\ref{fig:aspherity} extends the
range of $\gamma$ in Fig.~\ref{fig4} by six orders of magnitude.

%Fig 7

\subsubsection{Icosahedral spherical harmonics}
\label{sec:Q}

\begin{figure}
\centerline{\includegraphics[width=\linewidth]{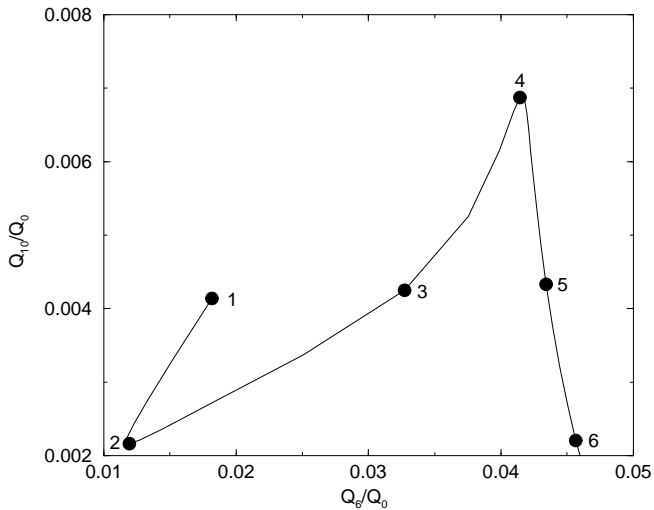}}
\caption{\label{fig:Q}
Combination of spherical harmonics (see text), which fall on a
universal curve for large enough shells.  Points for triangulations
defined by $(h,0), h = 5,6, \ldots,25$ were used to construct the
solid curve.  For $h < 5$ deviations due to discreteness become large
and those points have been omitted for clarity.  Points labeled 1--6
correspond to $\gamma \approx 0.5, 30, 1000, 15 000, 8 \times 10^6$
and $2.5 \times 10^8$, respectively.  }
\end{figure}

We first expand the radial density of points on the surface in
spherical harmonics,
\begin{equation}
	R(\theta,\phi) = \sum_{l,m} Q_{lm} Y_{lm}(\theta,\phi),
\end{equation}
where the density $R(\theta,\phi)$ is defined by
\begin{equation}
  R(\theta,\phi) = \sum_j R_j \delta(\phi - \phi_j) 
         \delta(\cos \theta - \cos \theta_j),
\end{equation}
and $(R_j, \theta_j, \phi_j)$ represents the polar coordinates of
vertex $j$.
From the coefficients $Q_{lm}$, we form the rotation-invariant
combinations
\begin{equation}
	Q_l^2 = \f{4\pi}{2l+1} \sum_{m=-l}^l \left| Q_{lm} \right|^2.
\end{equation}
For a shape of icosahedral symmetry the $Q_l$'s are non-zero only for
$l=0,6,10, \ldots$~\cite{Steinhardt}.  We plot, in Fig.~\ref{fig:Q},
$Q_{10}/Q_0$ vs $Q_6/Q_0$, which for large shells should fall on a
universal curve parameterized by $YR^2/\kappa$.
Note that {\em any} point in the $(Q_6/Q_0,Q_{10}/Q_0)$-plane would be
consistent with an icosahedral symmetry.  Continuum elastic theory,
however, predicts a {\em universal} set of functions $Q_l(\gamma)$
parametrized only by the Foppl-von K{\'a}rm{\'a}n number.
The buckling transition occurs between the points labeled 2--4 in
Fig.~\ref{fig:Q}, while the crossover to the ridge scaling for very
large $\gamma$ happens between points 4--6.

%Fig 8

\subsubsection{Curvature}

\begin{figure}
\centerline{\includegraphics[width=\linewidth]{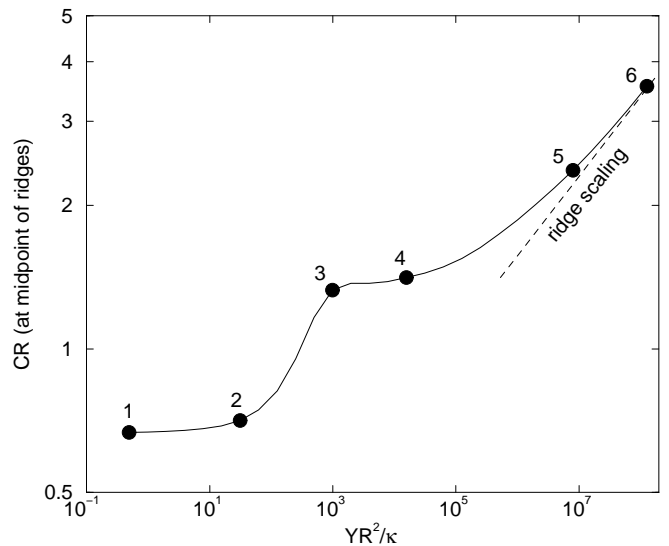}}
\caption{
\label{fig:curvature}
Curvature at the midpoints of the ridges.  The points labeled 1--6 are
for the same values of $\gamma$ as in Fig.~\ref{fig:Q}.  The dashed
line shows the asymptotic scaling behavior $\sim \gamma^{1/6}$.  }
\end{figure}

The curvature $C$ across the midpoints of the ridges connecting the
vertices of the icosahedra is plotted in Fig.~\ref{fig:curvature}.  As
$\gamma = YR^2/\kappa$ increases through the transition the ridges get
sharper and the shape becomes more faceted.  A perfect sphere would
have $C = 1/R$.  However, as seen in Fig.~\ref{fig:curvature}, $CR$
saturates to a slightly smaller value, $\simeq 0.7$, for very small
$\gamma$, implying that the shape is not perfectly spherical below the
buckling transition.  In fact, there is a weak tendency toward a
dodecahedral shape (which is the dual to the icosahedron).  The effect
is hardly visible (cf.\ Fig.~\ref{fig:aspherity}), however.
The data for large shells is well described by a scaling form
\begin{equation}
	C = R^{-1} F(Y R^2/\kappa),
\end{equation}
depending on the single parameter $\gamma$.
In the limit of large arguments $\gamma \to \infty$, we find
$F(\gamma) \to \gamma^{1/6}$, consistent with the scaling arguments of
Lobkovsky {\it et al.}~\cite{Lobkovsky,Lobkovskywitten}.
Note, however, that $\gamma$'s well in excess of $10^6$ are required
before one begins to see this asymptotic result of ``ridge scaling''.

%Fig 9

\section{Discussion}

\begin{figure}
\centerline{\includegraphics[width=\linewidth]{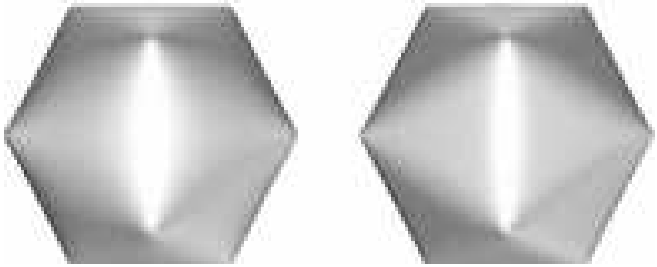}}
\caption{
\label{fig:two-shapes}
Two shapes for the large Foppl-von K{\'a}rm{\'a}n numbers $\gamma =
15600$ and $8 \times 10^6$ (points 4 and 5 in
Fig.~\ref{fig:Q}--\ref{fig:curvature}), illustrating the sharpening of
the ridges in the ridge scaling regime.}
\end{figure}

We have analyzed a model, based on the (highly nonlinear) physics of
thin elastic shells, which may be suitable for describing the shapes
of large viruses and of large vesicles with crystalline order in the
lipids.  Application of our results to vesicles~\cite{Blaurock,Zemb}
seems straightforward, once sufficiently precise freeze fracture or
confocal microscope images become available.
Figure~\ref{fig:two-shapes} illustrates two highly faceted shapes we
found for the large Foppl-von K{\'a}rm{\'a}n numbers or ``vK's'' which
might be relevant to the experiments of Ref.~\cite{Zemb}.  One
complication neglected here concerns possible phase separation of the
binary lipid mixtures studied by Dubois {\it et al.} in the vicinity
of the 12 disclinations~\cite{Zemb}.  It would be interesting to
investigate this effect, although a modest enrichment of one lipid
species near a disclination could be incorporated into a renormalized
core energy.  As discussed in Sec.~\ref{sec:II-B}, vK's in excess of
$10^6$ are required to see clearly the interesting scaling predictions
of Refs.~\cite{Lobkovsky} and \cite{Lobkovskywitten}.  In the
remainder of this section, we comment on the relevance of our work to
spherical viruses~\cite{Chiu,Baker,Caspar-Klug,Reddy}, where the vK's
are of order a few thousand or less.

%Fig 10

\begin{figure*}
\centerline{\includegraphics[width=\linewidth]{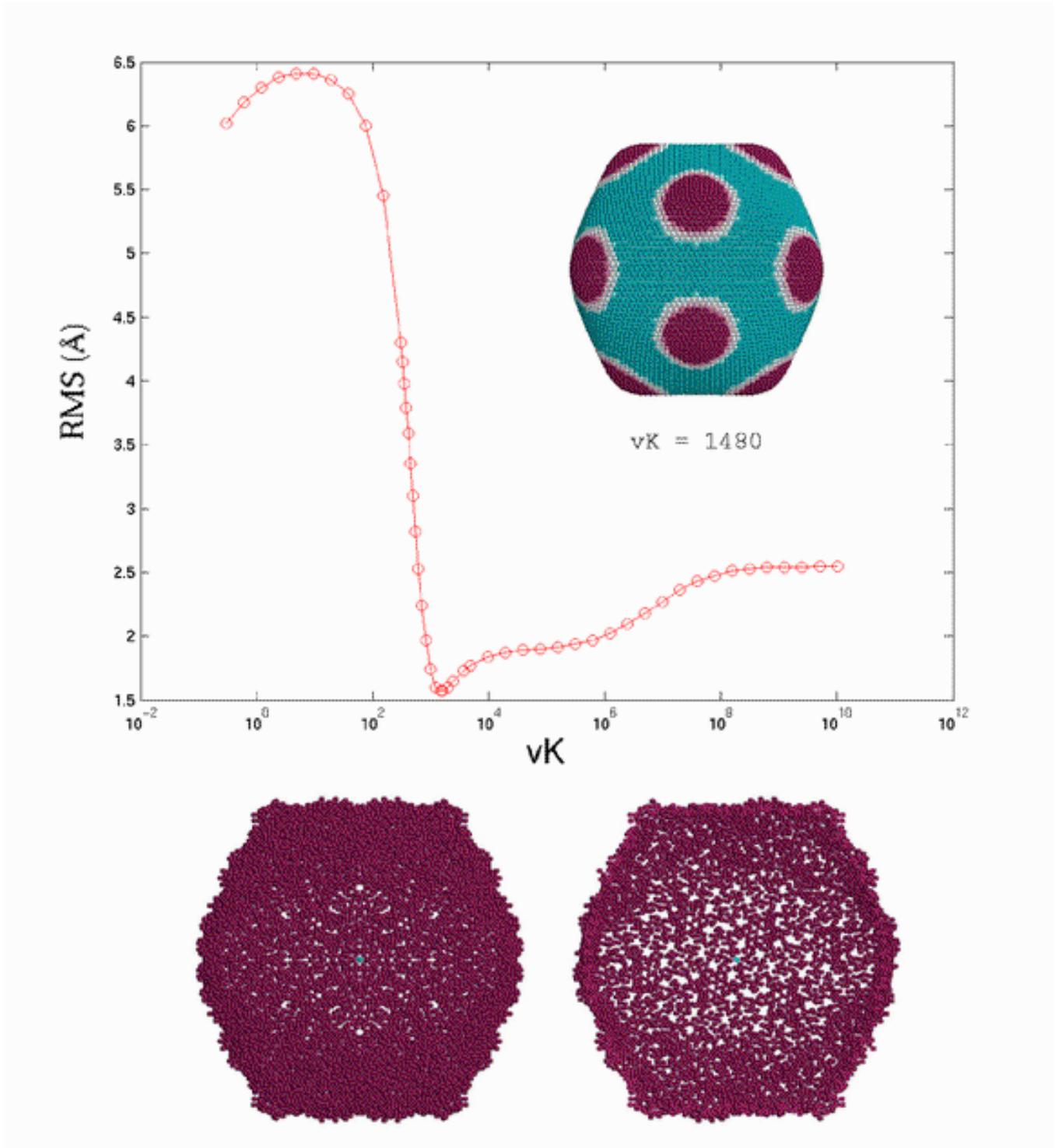}}
\caption{
\label{fig:11}
Real space fit to a virus structure.  The curve shows the
root-mean-square deviation from the experimental virus shell of
bacteriophage HK97 (full virus and cross section shown in the lower
part of the figure) and the theoretically calculated shape.  The best
fit occurs at the minimum for $\gamma \approx 1480$ and the
corresponding shape is depicted in the inset.  Shading indicates the
distance of the shell from the center. }
\end{figure*}

Ref.~\cite{Baker} compiles cryro-electron micrographs and other data
on approximately thirty different viruses, arranged in order of
increasing size.  These images highlight the trend that small viruses
are round and larger viruses are more faceted.  We view faceted
viruses as the result of 12 simultaneous buckling transitions,
centered on 12 disclinations, similar to the buckling of an {\em
isolated} disclination centered on a disk with open boundary
conditions~\cite{Seung-Nelson}.  The {\em spherical} packing of the
protein capsomers in viruses not only forces in twelve disclinations
(which we assume reside at the vertices of an
icosahedron)~\cite{Caspar-Klug}, but also breaks the up/down symmetry
of a disk with respect to the direction of buckling.  Hence, we expect
(and find numerically) that the sharp buckling transition with
increasing size in Ref.~\cite{Seung-Nelson} is smeared out.  As shown
in Fig.~\ref{fig:11} for bacteriophage HK97 (with 72 capsomers), good
one parameter fits in real space to the full three dimensional shape
of spherical viruses are possible.  Our best fit for Foppl-von
K{\'a}rm{\'a}n number of this mature form of HK97 is $\gamma =
YR^2/\kappa = 1480$, from which we can extract the ratio of the
Young's modulus $Y$ to the bending rigidity $\kappa$, given that the
virus diameter is $2R = 60$ nm~\cite{Baker}.
%Fig 11
The {\em precursor} capsid or ``prohead'' of HK97 is rather spherical,
in contrast to the larger, more faceted mature infectious virus shown
in Fig.~\ref{fig:11}~\cite{Conway}.  It seems likely that this virus
particle undergoes a buckling transition as it passes from the prohead
to its mature infectious form.
Indeed, the prohead shell is wrinkled or corrugated relative to the
mature form~\cite{Conway}.  It seems reasonable to regard this
transformation as a change in the effective thickness $d$ of the viral
shell: $d'$ of the prohead goes to $d < d'$ in the mature form.  As
discussed in Sec.~\ref{sec:II-A}, $\gamma = 12 (1-\nu_3^2) (R/d)^2$ if
we approximate the shell by a uniform isotropic elastic medium with
Poisson ratio $\nu_3$~\cite{Landau-Lifshitz}.  Since $2R$ {\em
increases} from 52 nm $\to$ 60 nm as $d$ decreases, it is plausible
that $\gamma \sim (R/d)^2$ rises and that the transformation from
prohead $\to$ head is accompanied by a buckling transition.

%Fig 12
\begin{figure*}
\centerline{\includegraphics[width=\linewidth]{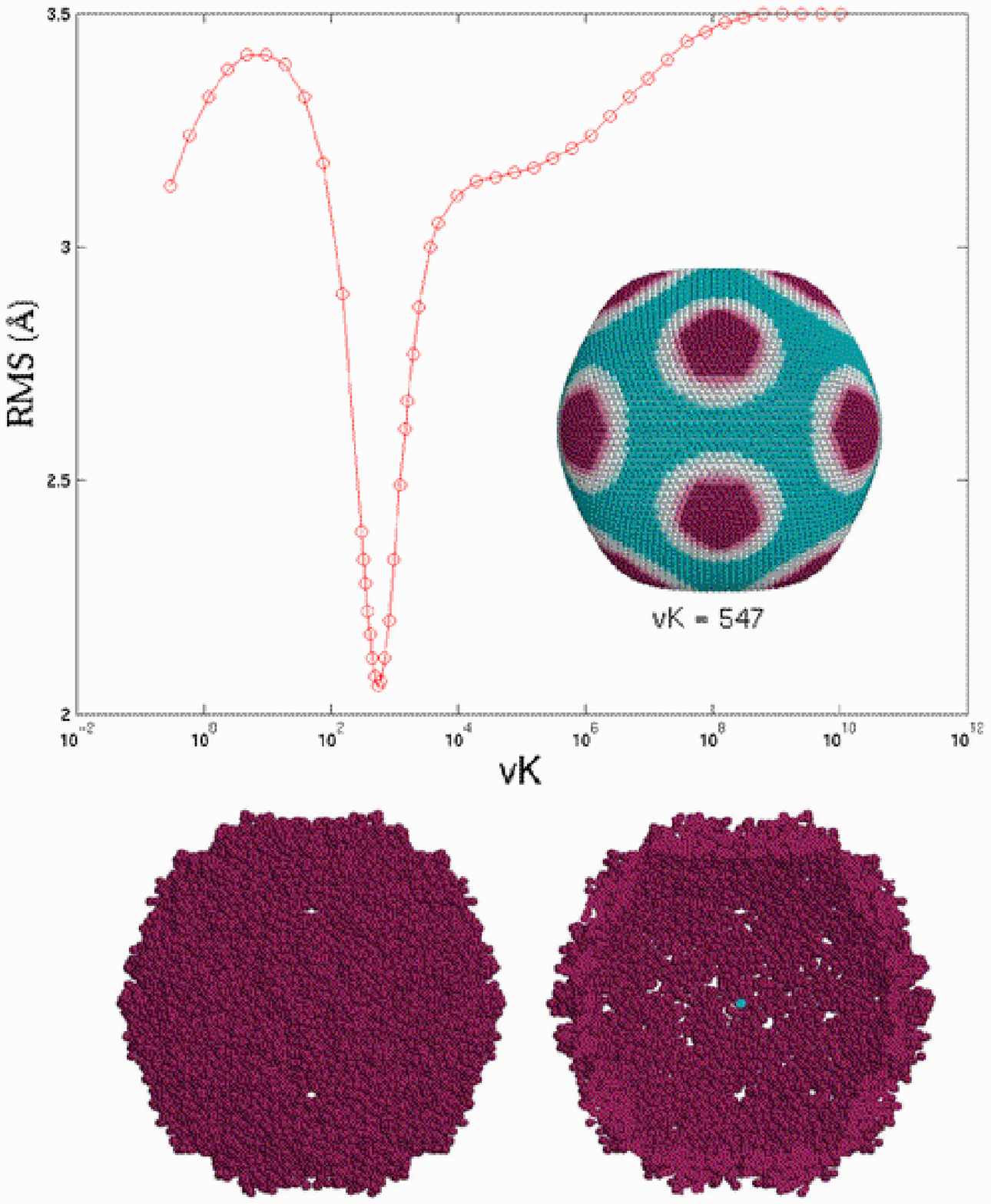}}
\caption{
\label{fig:12}
Real space fit to Yeast virus L-A with inset and actual virus as
described in Fig.~\ref{fig:11}.  The best fit occurs for $\gamma =
547$.  }
\end{figure*}

Figure \ref{fig:12} shows a similar fit to the yeast L-A virus, which
yields $\gamma = YR^2/\kappa = 547$.  The diameter of the yeast virus
is $2R = 43$ nm, which leads to the conclusion that
$[Y/\kappa]_\mathrm{yeast} = 1.24$ nm$^{-2}$.  Note that
$[Y/\kappa]_\mathrm{HK97} = 1.64$ nm$^{-2}$, consistent with the
arguments given in the Introduction that spherical viruses have
roughly similar elastic constants.

Our models for virus shells are based on two important assumptions.
The first is the neglect of a spontaneous curvature
term~\cite{Leibler} in bending energies such as Eq.~\ref{bend}.  Such
a term might be significant if the viral building blocks had a
pronounced conical shape similar to, say, surfactant molecules in a
micelle~\cite{Leibler,Safran} or laboratory cork stoppers.  Several
lines of evidence suggest that neglect of this term might be
justified.  Although certain virus scaffolding proteins (which can act
as templates for early phases of construction) do have a conical
shape, these are discarded in the mature icosahedral viruses of
interest to us here~\cite{Chiu}.  It is hard to see a strong mechanism
for precisely-defined hinge angles in very large viruses composed of
many capsomers~\cite{Reddy2,Sorger}.  In vitro assembly experiments on
the polyoma virus do produce spherical aggregates with 12, 24 and 72
pentameric units depending on conditions of pH, calcium concentration,
etc~\cite{Salunke}, which could be accounted for by a spontaneous
curvature term.  Although we neglect spontaneous curvature here, it is
certainly possible that the physics of small viruses (or the
scaffolding proteins themselves) are influenced by such an energy, as
has been explored by Bruinsma, Rudnick and Gelbart~\cite{Bruinsma}.

In general our work is more likely to be applicable to {\em large}
viruses, for which simple continuum models can be justified.
Icosahedral viruses are usually composed of a combination of 5- and
6-fold symmetric packing units, with the twelve 5-fold units centered
on the vertices of an icosahedron (the polyoma virus SV40~\cite{Baker}
with its 72 identical pentamers is an exception).  Because the strain
energies which lead to buckling extend far from the disclinations
which produce them~\cite{Seung-Nelson}, we would expect our results
for large viruses to be insensitive to differences in the shapes of
packing elements.  Special packing elements at the twelve 5-fold sites
could be incorporated into a disclination core energy.

A second key assumption is our neglect of the osmotic pressure due to
the confined DNA or RNA package of the virus~\cite{Smith}.  Here we
can appeal to an experiment.  Earnshaw and Harrison~\cite{Earnshaw}
have compared the structure of phage lambda (P22) with its full
complement of DNA to the structure of lambda mutants containing only
78\% of the native DNA.  Although changes in the details of DNA
packing can be detected, the protein shell itself is unchanged.  Thus,
due either to DNA condensation or an exceptionally strong shell, the
osmotic pressure of the DNA is insufficient to change the shape.  Of
course, the nucleic acid content of a virus could nevertheless play an
important role in shell assembly~\cite{Rudnick}.

\begin{acknowledgments}

It is a pleasure to acknowledge valuable advice and conversations with
S.\ Harrison, B.\ Shraiman, J.\ Johnson, S.\ Burley, R.\ Bruinsma, P.\
Lenz, J. Israelachvilli, and T.\ Zemb.  Work by DRN and JL was
supported in part by the National Science Foundation, through grant
DMR-0231631 and through the Harvard Material Science and Engineering
Center through grant DMR-0213805.  DRN would like to acknowledge the
hospitality and support of the Center for Studies in Physics and
Biology at Rockefeller University in the fall of 2000, where this work
was begun.  JL acknowledges support from the Swedish Research Council
(VR).  LM acknowledges support from the Alfred P.\ Sloan foundation.

\end{acknowledgments}


\begin{thebibliography}{99}

\bibitem{Chiu}
Structural Biology of Viruses, edited by W. Chiu,
R. M. Burnett and R. L. Garcea
(Oxford Univeristy Press, New York 1997).

\bibitem{Harrison-91}
S. C. Harrison,
``What do Viruses Look Like?'',
Harvey Lectures 85, 127 (1991).

\bibitem{Baker}
T. S. Baker, N. H. Olson and S. D. Fuller,
``Adding the Third Dimension to Virus Life Cycles'',
Microbiology and Molecular Biology Reviews, 63, 862 (1999).

\bibitem{Smith} D. E. Smith, S. J. Tans, S. B. Smith, S. Grimes,
D. L. Anderson and C. Bustamante,
``The bacteriophage $\phi 29$ portal motor can package DNA against
a large internal force'',
Nature 413, 748 (2001).

\bibitem{Landau-Lifshitz} L.D. Landau and I.M. Lifshitz,
\emph{Theory of Elasticity} 
(Pergamon, New York, 1975).

\bibitem{Crick} F. Crick and and J. D. Watson,
``The Structure of Small Viruses'', 
Nature {\bf 177}, 473 (1956).

\bibitem{Caspar-Klug} D. Caspar and A. Klug,
``Physical Principles in the Construction of Regular Viruses'',
Cold Spring Harbor Symposium on Quantitative Biology, {\bf 27}, 1 (1962).

\bibitem{Coxeter} H.M.S. Coxeter,
\emph{Introduction to Geometry} (Wiley and Sons, Inc., 
New York 1969), Chapts. 19--21.

\bibitem{Reddy}
V. Reddy, P. Natarajan, B. Okerberg, K. Li, K. Damodaran,
M. R. Brooks III, J. Johnson,
``Virus Particle Explorer (VIPER), a Website for Virus Capsid Structures and 
Their Computational Analyses'',
Journal of Virology {\bf 75}, 11943 (2001);
%
P. L. Stewart, R. M. Burnett and S. D. Fuller,
``Image reconstruction reveals the complex molecular organization of
adenovirus'', 
Cell {\bf 67}, 145-167 (1991).

%\bibitem{Harrison-2} S. Harrison, private communication.

\bibitem{Murzin}
A. G. Murzin, S. E. Brenner, T. Hubbard, C. Chothia,
``SCOP: a structural classification of proteins database for the
investigation of sequences and structures'',
J. Mol. Biol. {\bf 247}, 536-540 (1995),
\url{http://scop.mrc-lmb.cam.ac.uk/scop/data/scop.b.c.bd.b.html}.

\bibitem{Carrion-Vazquez}
M. Carrion-Vazquez, A. F. Oberhauser, T. E. Fisher, P. E. Marszalek,
H. Li, J. M. Fernandez,
``Mechanical design of proteins studied by single-molecule force
spectroscopy and protein engineering'',
Prog. Biophys. Mol. Biol. {\bf 74}, 63-91 (2000).

\bibitem{Kellermayer}
M. S. Kellermayer, C. Bustamante, H. L. Granzier,
``Mechanics and structure of titin oligomers explored with atomic force
microscopy'',
Biochim. Biophys. Acta. {\bf 1604}, 105 (2003).

\bibitem{Seung-Nelson} S. Seung and D. R. Nelson,
``Defects in Flexible Membranes with Crystalline Order'',
Phys. Rev. A{\bf 38}, 1005 (1988).

\bibitem{Nelson-Peliti} D. R. Nelson and L. Peliti,
``Fluctuations in Membranes with Crystalline and Hexatic Order'',
Journal de Physique {\bf 48}, 1085 (1987), and references therein.

\bibitem{Bowick} M.J. Bowick, D.R. Nelson, and A. Travesset, 
``Interacting Topological Defects in Frozen Topographics,''
Phys. Rev. B{\bf 62}, 8738 (2000);
(\eprint{arXiv:cond-mat/9911379});
See also A. R. Bausch, M. J. Bowick, A. Cacciuto, A. D. Dinsmore,
M. F. Hsu, D. R. Nelson, M. G. Nikolaides, A. Travesset, and
D. A. Weitz,
``Grain Boundary Scars and Spherical Crystallography'',
Science {\bf 299}, 1716 (2003).

\bibitem{Lobkovsky}
A. E. Lobkovsky,
``Boundary layer analysis of the ridge singularity in a thin plate'',
Phys. Rev. E {\bf 53}, 3750 (1996).

\bibitem{Lobkovskywitten} A. E. Lobkovsky and T. A. Witten,
``Properties of Ridges in Elastic Membranes'',
Phys. Rev. E{\bf 55}, 1577 (1997).

\bibitem{Cerda} E. Cerda and L. Mahadevan
``Conical Surfaces and Crescent Singularities in Crumpled Sheets'',
Phys. Rev. Lett. {\bf 80}, 2358 (1998).

\bibitem{Blaurock} A. E. Blaurock and R. C. Gamble,
``Small Lecithin Vesicles Appear to be Faceted below the Thermal
Phase Transition'',
J. Membrane Biol. {\bf 50}, 187 (1979);
%\bibitem{Sackmann}
E. Sackmann and R. Lipowsky,
``Handbook of Biological Physics'', Vol. 1B, ed. A. J. Hoff,
(North-Holland, Amsterdam, 1995).

\bibitem{Zemb} M. Dubois, B. Dem\'e, T,. Gulik-Krzywicki, J.-C. Dedieu, 
C. Vantrin, S. Desert, E. Perz, and T. Zemb, 
``Self-Assembly of Regular Hollow Icosahedra in Salt-free Catanionic
Solutions.'',
Nature {\bf 411}, 672 (2001).

\bibitem{Kantor-Nelson} Y. Kantor and D. R. Nelson,
``Phase Transitions in Flexible Polymeric Surfaces'',
Phys.  Rev. A{\bf 36}, 4020 (1987).

\bibitem{Yan} X. Yan, N. H. Olson, J. L. Van Etten, M. Bergoin,
M. G. Rossmann, and T. S. Baker, 
``Structure and Assembly of Large Lipid-Containing dsDNA Viruses'',
Nature Structural Biology {\bf 7}, 101 (2000).

\bibitem{Steinhardt} P. Steinhardt, D. R. Nelson and M. Ronchetti,
``Bond orientational order in liquids and glasses'',
Phys, Rev. B {\bf 28}, 784 (1983).

\bibitem{Seifert} For an analogous exploration of the shapes of liquid
  membranes with a spherical topology, see U. Seifert,
 ``Configurations of  Fluid Membranes and Vesicles'',
Advances in Physics {\bf 46}, 13 (1997);
See also 
%\bibitem{Noguchi}
H. Noguchi,
 ``Polyhedral vesicles: A Brownian dynamics simulation'',
Phys. Rev. E {\bf 67}, 041901 (2003).

\bibitem{Nelson-book} See, e.g., D.R. Nelson,
\emph{Defects and Geometry in Condensed Matter Physics}
[Cambridge University Press, Cambridge, 2002].

\bibitem{Leibler}
See, e.g., S. Leibler in
``Statistical Mechanics of Membranes and Interfaces'', edited by
D. R. Nelson, T. Piran and S. Weinberg
(World Scientific, Singapore 1989).

\bibitem{Conway}
J. F. Conway, W. R. Wikoff, N. Cheng, R. L. Duda, R. W. Hendrix,
J. E. Johnson, and A. C. Steven,
 ``Virus Maturation Involving Large Subunit Rotations and Local Refolding'',
Science {\bf 292}, 744 (2001).

\bibitem{Safran}
S. A. Safran,
``Statistical Thermodynamics of Surfaces, Interfaces and Membranes'',
(Addison-Wesley, New York 1994).

\bibitem{Reddy2}
V. S. Reddy, H. A.  Giesing, R. T. Morton, A. Kumar, C. B. Post,
C. L. Brooks III, and J. E. Johnson,
 ``Energetics of quasiequivalence: computational analysis of protein-
 protein interactions in icosahedral viruses'',
Biophys. {\bf 74}, 546 (1998).

\bibitem{Sorger}
P. K. Sorger, P. G. Stockley, S. C. Harrison,
 ``Structure and Assembly of Turnip Crinkle Virus .2. Mechanism of
 Reassembly Invitro'',
J. Mol. Biol. {\bf 191}, 639 (1986).

\bibitem{Salunke}
D. M. Salunke, D. L. Caspar, and R. L. Garcea,
 ``Polymorphism in the assembly of polyomavirus capsid protein VP1'',
Biophys. {\bf 56}, 887 (1989).

\bibitem{Bruinsma}
R. F. Bruinsma, W. M. Gelbart, D. Reguera, J. Rudnick and R. Zandi,
 ``Viral self-assembly as a thermodynamic process'',
Phys. Rev. Lett. {\bf 90}, 248101 (2003) (\eprint{cond-mat/0211390}).

\bibitem{Earnshaw}
W. C. Earnshaw and S. C. Harrison,
 ``DNA arrangement in Isometric Phage Heads'',
Nature {\bf 268}, 598 (1977).

\bibitem{Rudnick}
J. Rudnick and R. F. Bruinsma,
 ``Icosahedral packing of RNA viral genomes'',
\eprint{cond-mat/0301305}, and references therein.

\end{thebibliography}
\end{document}